\documentstyle[12pt,fullpage,subeqn,subeqnarray,epsf,epsfig]{article}
\def\be{\begin{equation}}
\def\ee{\end{equation}}
\def\beq{\begin{eqnarray}}
\def\eeq{\end{eqnarray}}
\def\lsim{\:\raisebox{-0.5ex}{$\stackrel{\textstyle<}{\sim}$}\:}
\def\gsim{\:\raisebox{-0.5ex}{$\stackrel{\textstyle>}{\sim}$}\:} 
\begin{document}
\begin{flushright}
TIFR/TH/99-11
\end{flushright}
\bigskip
\begin{center}
{\Large{\bf Neutrino Mass and Oscillation : An Introductory Review$^*$}} \\[1cm]
D.P. Roy \\
Tata Institute of Fundamental Research \\
Homi Bhabha Road, Colaba \\
Mumbai - 400 005, INDIA \\[2cm]
\underbar{\bf Abstract}
\end{center}
\bigskip

After a brief introduction to neutrino mass via the see-saw model I
discuss neutrino mixing and oscillation, first in vacuum and then its
matter enhancement.  Then the solar and atmospheric neutrino
oscillation data are briefly reviewed.  Finally I discuss the problem
of reconciling hierarchical neutrino masses with at least one large
mixing, as implied by these data.  A minimal see-saw model for
reconciling the two is discussed.

\vspace{7cm}

\hrule width 5cm

\smallskip

\noindent $^*$ Invited talk at the Symp. on Frontiers of Fundamental Physics,
Hyderabad, 30 December 98 - 1 January 99 and the Discussion Meeting on Recent
Developments in Neutrino Physics, Ahmedabad, 2-4 February 99.

\newpage

\noindent {\Large{\bf Neutrino Mass : See-saw Model}}
\medskip

The fermion masses are represented by the combination of Dirac spinors
\be
m \bar u_L u_R,
\label{one}
\ee
while
\be
\bar u_L u_L = \bar u_R u_R = 0.
\label{two}
\ee
It follows from the basic anticommutation relation of Dirac $\gamma$
matrices that the bar of a left-handed projection operator is
right-handed, so that the product of the opposite projection operators
vanish identically.  In the Standard Model (SM) the left-handed
fermions occur in $SU(2)$ doublets and the right-handed ones in
singlets except that there is no right-handed neutrino, i.e.
\be
\left(\matrix{u_i \cr d_i}\right)_L, u_{iR}, d_{iR}; \
\left(\matrix{\ell_i \cr \nu_i}\right)_L, \ell_{iR}.
\label{three}
\ee
Here the particle labels represent the corresponding spinors and $i$
is the generation index.  Thus we can not have direct mass terms even
for quarks and charged leptons, since terms like $m \bar\ell_L \ell_R$
are not gauge invariant.  However they can acquire mass by absorbing
the Higgs scalar $\phi$, which is a $SU(2)$ doublet.  Thus the Yukawa
interaction, $f\phi \bar\ell_L \ell_R$, is gauge invariant.  As the
gauge symmetry is spontaneously broken, $\phi$ acquires a vacuum
expectation value and correspondingly the fermion acquires a mass,
i.e. 
\be
f \phi \bar\ell_L \ell_R \rightarrow \underbrace{f\langle \phi
\rangle}_m \bar\ell_L \ell_R.
\label{four}
\ee
However this is not possible for the neutrino, since it does not have
a right-handed component in the SM.

The simplest and most popular way of giving mass to the neutrino is
via the see-saw model [1].  It assumes each neutrino to have a
right-handed singlet component $N_R$ like the other fermions.  Unlike
the latter however the $N_R$ has a unique property.  Its gauge charges
corresponding to the SM gauge groups 
\be
SU(3)_C \times SU(2) \times U(1)_Y
\label{five}
\ee
are all zero --- it carries no colour, isospin or hypercharge.  It is
called a Majorana particle, since the particle and antiparticle have
the same gauge charges --- minus zero is zero.  Of course they have
opposite lepton numbers, which is however not a gauge quantum number
and hence need not be conserved.  Consequently one can have a direct
mass term coupling the right-handed singlet neutrino with its
left-handed antiparticle,
\be
M \overline{N^C_L} N_R,
\label{six}
\ee
which is called Majorana mass.  This can be very large since it does
not break any gauge symmetry.  In addition one can have a Dirac mass
term like (\ref{four}), i.e.
\be
m \bar\nu_L N_R.
\label{seven}
\ee
Diagonalising the resulting mass-matrix in the $\nu_L - N_R$ basis
induces a tiny mass for the left-handed neutrino, i.e.
\be
\left(\matrix{0 & m \cr m & M}\right) \rightarrow \left(\matrix{m^2/M
& 0 \cr 0 & M}\right).
\label{eight}
\ee
The larger the Majorana mass of $N_R$ the smaller will be the
left-handed neutrino mass induced by it, since $m_\nu = m^2/M$.
Therefore it is called see-saw model.  It is an ingenuous model that
solves two puzzles at one stroke.  The large Majorana mass $M$
banishes the right-handed neutrino beyond observation.  It also makes
the left-handed neutrino mass tiny compared to the other fermion
masses, which are characterised by the Dirac mass $m$.

The way the Majorana mass (\ref{six}) was introduced above as a direct
mass term was rather adhoc.  This was remedied in [2] by assuming an
extension of the SM gauge group (\ref{five}) by a
\be
U(1)_{B-L},
\label{nine}
\ee
whose gauge charge corresponds to the difference of Baryon and Lepton
numbers.  In this case the requirement of anomaly cancellation implies
the existence of 3 right-handed singlet neutrinos as in the case of
the other fermions.  This will ensure vector coupling of the
$U(1)_{B-L}$ current as the axial parts cancel between the left and
right handed fermions.  Then one can easily see that the remaining
axial anomalies cancel one by one.  Therefore the $U(1)_{B-L}$ can be
treated as a gauge symmetry along with the other symmetry groups of
the SM.  One assumes spontaneous breaking of this gauge symmetry at a
high mass scale via a Higgs scalar $\chi$, carrying 2 units of lepton
number (i.e. $B-L = 2$).  The coupling of this Higgs scalar leads to
the Majorana mass
\be
f \chi \overline{N^C_L} N_R \rightarrow \underbrace{f\langle \chi
\rangle}_M \overline{N^c_L} N_R,
\label{ten}
\ee
analogous to the Dirac mass (\ref{four}).  This is a left-right
symmetric model, which can be embeded in $SO(10)$ GUT.  Indeed all the
fermions of one generation, including $N_R$, can be naturally
accommodated in the 16 dimensional representation of $SO(10)$.  In the
$SU(5)$ GUT on the other hand the $10 + 5$ dimensional representation
can naturally accommodate all the SM fermions of one generation.  In
this sense the $SU(5)$ GUT is more appropriate for the SM.  But even
in this case one can add a $N_R$ as a $SU(5)$ singlet if needed.  It
may be noted here that in $SU(5)$ GUT one has the flexibility of
adding any number of right-handed singlet neutrinos --- not
necessarily three.  In fact we shall see later that a minimal solution
to the atmospheric and solar neutrino oscillation data requires only
two right handed singlet neutrinos.
\bigskip

\noindent {\Large{\bf Neutrino Mixing and Oscillation (Vacuum) :}}
\medskip

If the neutrinos have nonzero mass, there will in general be mixing
between the neutrino species as in the case of quarks.  For most
practical applications it is adequate to consider mixing between a
pair of neutrino species, e.g.
\be
\left(\matrix{\nu_e \cr \nu_\mu}\right) = \left(\matrix{\cos\theta &
\sin\theta \cr -\sin\theta & \cos\theta}\right) \left(\matrix{\nu_1
\cr \nu_2}\right),
\label{eleven}
\ee
where $\nu_{1,2}$ are the mass eigen states with eigen values
$m_{1,2}$.  Each of them propagates with a distinct phase factor,
characterised by its mass.  To see this we make the extreme
relativistic approximation, appropriate for the tiny neutrino masses,
i.e. 
\be
E_{1,2} \simeq p + m^2_{1,2}/2p \ {\rm and} \ t \simeq \ell.
\label{twelve}
\ee
Then the phase factors for the free particle wave functions of
$\nu_{1,2}$ are given by
\be
e^{-i(E_{1,2} t - p\ell)} \simeq e^{-{im^2_{1,2} \ell \over 2E}}.
\label{thirteen}
\ee
This automatically leads to neutrino oscillation [3] as we see below.

Let us consider a beam of $\nu_e$ produced at the origin.  It can be
split into the $\nu_1$ and $\nu_2$ components, each propagating with
its own phase, i.e.
\be
\nu_e \rightarrow \nu_1 \cos\theta e^{-{im^2_1 \ell \over 2E}} +
\nu_2 \sin\theta e^{-{im^2_2 \ell \over 2E}}.
\label{fourteen}
\ee
To find the $\nu_\mu$ component in this beam after a certain distance
$\ell$, we split $\nu_{1,2}$ back to $\nu_{e,\mu}$, i.e.
\beq
\nu_e \rightarrow (\cos\theta \nu_e \!\!&-&\!\! \sin\theta \nu_\mu)\cos\theta
e^{-im^2_1 \ell \over 2E} \nonumber \\[2mm] \!\!&+&\!\! (\sin\theta \nu_e +
\cos\theta \nu_\mu) \sin\theta e^{-im^2_2\ell \over 2E}.
\label{fifteen}
\eeq
Clearly the coefficient of the $\nu_\mu$ term is nonzero.  The square
of this co-efficient represents the oscillation probability, i.e.
\beq
P_{\nu_e \rightarrow \nu_\mu} (\ell) &=& \left|\cos\theta\sin\theta
\left(-e^{-im^2_1\ell \over 2E} + e^{-im^2_2\ell \over
2E}\right)\right|^2 \nonumber \\[2mm]
&=& \sin^2 2\theta sin^2 {\delta m^2 \ell \over 4E},
\label{sixteen}
\eeq
where $\delta m^2 = m^2_1 - m^2_2$ and the two factors represent the
amplitude and the phase of oscillation.  Converting to more convenient
units, one gets
\beq
P_{\nu_e \rightarrow \nu_\mu} (\ell) &=& \sin^2 2\theta \sin^2
\left(1.3 \underbrace{\delta m^2}_{eV^2} \cdot \underbrace{\ell}_m /
\underbrace{E}_{\rm MeV}\right) \nonumber \\[2mm]
&=& \sin^2 2\theta \sin^2 (\ell \pi/\lambda),
\label{seventeen}
\eeq
where $\lambda$ represents the wave length of oscillation.  Note that
the oscillation probability reaches maxima at odd multiples of 
\be
\lambda/2 (m) = 1.25 E ({\rm MeV})/\delta m^2 ({\rm eV}^2),
\label{eighteen}
\ee
while it vanishes at the even multiples.  Thus for large mixing angle,
$\sin 2\theta \simeq 1$, one can identify three distinct distance
scales of oscillation, i.e.
\beq
\ell & & \ll \lambda/2 \simeq \lambda/2 \gg \lambda/2 \nonumber\\[2mm]
{\displaystyle P_{\nu_e \rightarrow \nu_\mu} (\ell) \atop (1 -
P_{\nu_e \rightarrow \nu_e} (\ell))} &=& \hspace{.5cm} 0 \hspace{1cm}
1 \hspace{1cm} 1/2, 
\label{ninteen}
\eeq
where the last quantity comes from averaging $P$.

It is clear from the above discussion that assuming large mixing one
can see the effect of neutrino oscillation for
\be
\ell (m) \gsim 1.25 E({\rm MeV})/\delta m^2 ({\rm eV}^2),
\label{twenty}
\ee
and the same relation holds if we replace the units of distance and
energy by Km and GeV respectively.  Let us illustrate this with some
real life examples.

\begin{center}
Table I. Sensitivity of different types of neutrino experiments \\ to the
scales of neutrino mass
\end{center}
\[
\begin{tabular}{|c|c|c|c|}
\hline
$\nu$ Source & Energy $(E)$ & Dist. $(\ell)$ & $\delta m^2$ \\
\hline
Reactor & $\sim$ MeV & $10^2 \ m$ & $\gsim 10^{-2}$ \ eV$^2$ \\
Sun & -''- & $10^{11} \ m$ & $\gsim 10^{-11} \ {\rm eV}^2$ \\
\hline
Accelerator & $\sim$ GeV & Km & $\gsim \ {\rm eV}^2$ \\
Atmosphere & -''- & $10^4$ Km & $\gsim \ 10^{-4} \ {\rm eV}^2$ \\
\hline
\end{tabular}
\]

\noindent Table I summarises the typical energy and distance scales
for the four different types of neutrino experiments.  The reactor and
solar neutrinos, arising from nuclear reactions, have energies in the
MeV range.  The typical distance travelled is $\sim 10^2$ meters in the
former case and $1 \ {\rm AU} \simeq 10^{11} \ m$ in the latter.  Thus
from (\ref{twenty}) the reactor neutrino oscillation experiments are
sensitive to $\delta m^2 \gsim 10^{-2} \ {\rm eV}^2$ while the solar neutrino
oscillation is sensitive down to $10^{-11} \ {\rm eV}^2$.  The
accelerator and atmospheric neutrinos, arising from pion decay in
flight, have typical energies in the GeV range.   The typical distance
scale for short base line accelerator neutrino experiments is $\sim 1$
Km, while it is $\sim 10^4$ Km for atmospheric neutrinos traversing
through the diameter of the earth.  Thus the accelerator neutrino
oscillation experiments are sensitive to $\delta m^2 \gsim {\rm
eV}^2$, while atmospheric neutrino oscillation is sensitive down to
$10^{-4} \ {\rm eV}^2$.

Of course the above realisation is not new.  Soon after the discovery
of neutrino, it was observed by Pontecorvo [3] that the solar neutrino
oscillation experiment can probe for neutrino mass down to a small
fraction of an eV, if there is significant mixing between the neutrino
species.  Indeed this provided the main motivation for starting the
solar neutrino experiment in the late sixties.  Later on it was
realised that the solar neutrinos can show large oscillation even for
small mixing between the neutrino species [4-6], as we see below.
\bigskip

\noindent {\Large{\bf Matter Enhancement (Resonce Oscillation) :}}
\medskip

This phenomenon is also known as the MSW effect, as it was
systematically worked out by Mikheyev and Smirnov [4] following the
original suggestion of Wolfenstein [5].  This is analogous to the
effect of the medium on the propagation of light, which imparts an
induced mass to the photon, resulting in the index of refraction.
Similarly the neutrino, propagating through the sun, acquires an
induced mass.

The electrons in the solar medium has charged current interaction with
$\nu_e$,
\be
\nu_e e \ {\buildrel W \over \longrightarrow} \ e \nu_e,
\label{twentyone}
\ee
but not with $\nu_\mu$ or $\nu_\tau$.  The resulting interaction
energy is given by
\be
H_{\rm int} = \sqrt{2} G_F N_e,
\label{twentytwo}
\ee
where $G_F$ and $N_e$ are the Fermi coupling and the electron density
in the solar medium.  The corresponding neutral current interactions
are identical for all neutrino species and hence have no net effect on
their propagation.  To see the effect of the charged current
interaction let us consider the wave equation for $\nu_e$ and
$\nu_\mu$ as in the last section, but now including the interaction
energy (\ref{twentytwo}).  We have
\be
-{id \over dt} \left(\matrix{\nu_e \cr \nu_\mu}\right) = \left(p +
{M^2 + 2p H_{\rm int} \over 2p}\right) \left(\matrix{\nu_e \cr
\nu_\mu}\right). 
\label{twentythree}
\ee
Defining the quantity
\be
M^{\prime 2} = M^2 + 2p H_{\rm int} \simeq M^2 + 2E H_{\rm int},
\label{twentyfour}
\ee
as an effective mass or energy, we see that
\beq
M^{\prime 2} &=& \left(\matrix{c & s \cr -s & c}\right)
\left(\matrix{m^2_1 & 0 \cr 0 & m^2_2}\right) \left(\matrix{c & -s \cr
s & c}\right) + \left(\matrix{2\sqrt{2} E G_F N_e & 0 \cr 0 &
0}\right) \nonumber \\[2mm] 
&=& \left(\matrix{c^2 m^2_1 + s^2 m^2_2 + 2\sqrt{2} E G_F N_e &
sc\delta m^2 \cr sc\delta m^2 & c^2 m^2_2 + s^2 m^2_1}\right),
\label{twentyfive}
\eeq
where $s,c$ denote $\sin\theta$, $\cos\theta$.  The 1st term
represents the squared neutrino masses, rotated into the flavour
basis, while the 2nd term represents the interaction energy
(\ref{twentytwo}). 

Let us consider small mixing angle, $s \ll 1$, so that the eigen
states of (\ref{twentyfive}) correspond approximately to the flavour
eigen states $\nu_e$ and $\nu_\mu$.  The corresponding eigen values
are shown against the electron density in Fig. 1 for the case $m_2 >
m_1$.  They roughly correspond to the two diagonal elements of
(\ref{twentyfive}).  At the solar surface $N_e \rightarrow 0$, so that
the energy eigenvalues of $\nu_e$ and $\nu_\mu$ correspond to their
masses.  As one goes towards the solar core, however, the eigen value
of $\nu_e$ increases with $N_e$ and ultimately overtakes that of
$\nu_\mu$.  The two energy levels cross over at $M_{11}^{\prime 2} =
M^{\prime 2}_{22}$, i.e.
\be
N^c_e = {\delta m^2 \over 2\sqrt{2} G_F E} \cos 2\theta.
\label{twentysix}
\ee
Of course the finite off-diagonal elements ensure that the two energy
levels are separated at this point by a small but finite gap
\be
\Gamma = \delta m^2 \sin 2\theta,
\label{twentyseven}
\ee
which represents the width of the energy interval over which the level
crossing takes place.  
In analogy with vacuum oscillation, one can define an effective mixing
angle in matter as
\be
\tan 2\theta_M = {2M^{\prime 2}_{12} \over M^{\prime 2}_{22} -
M^{\prime 2}_{11}} = {\sin 2\theta \over \cos 2\theta - 2\sqrt{2} EG_F
N_e/\delta m^2}.
\label{twentyeight}
\ee
No matter how small the mixing angle $\theta$ the resonance condition
(\ref{twentysix}) ensures that $\theta_M \rightarrow 45^\circ$ at the
cross-over point.  This is why it is called matter enhanced (or
resonant) oscillation.

Physically speaking, the $\nu_e$ produced at the solar core has an
energy level higher than that of $\nu_\mu$ (Fig. 1).  It continues to occupy
this higher energy level as it emerges through the region of critical
density $N^c_e$, which means that it emerges out as $\nu_\mu$.  This
assumes of course that the transition probability between the two
energy levels at the critical point remains small, i.e.
\be
T \propto \gamma^{-1}_c \propto {2E \over \delta m^2} {\cos 2\theta
\over \sin^2 2\theta} \left({d N_e/d\ell \over N_e}\right)_c \ll 1.
\label{twentynine}
\ee
This is called the adiabatic condition.  One can show from
(\ref{eighteen}), (\ref{twentysix}) and (\ref{twentyseven}) that the
above expression represents the ratio of the oscillation wave-length
$\lambda$ to the distance over which the energy eigen value changes by
an amount $\Gamma$.  Thus the adiabatic conditon (\ref{twentynine})
requires the $N_e$ to change very slowly during the level crossing, so
that the resulting change 
in the energy over a distance $\lambda$ is $\ll \Gamma$ [6].  Such a
slow change of energy ensures that the wave function continuously
adjusts itself to one energy level (upper or lower) in
stead of jumping to the other.

Thus the twin conditions for the resonant conversion of $\nu_e
\rightarrow \nu_\mu$ inside the sun are $N_e^o > N^c_e$ and the
adiabatic condition (\ref{twentynine}).  Quantitively speaking the
conditions for $\geq 50$\% conversion of solar neutrino,
\be
\langle P_{\nu_e \rightarrow \nu_e}\rangle < 0.5
\label{thirty}
\ee
are
\be
{\delta m^2 \cos 2\theta \over 2\sqrt{2} G_F N^o_e} < E < {\pi \delta
m^2 \sin^2 2\theta \over 4\ell n 2 \cos 2\theta (N_e^{-1} d
N_e/d\ell)_c}.
\label{thirtyone}
\ee
The 1st inequality ensures resonant transition by requiring that the
core density is higher than the critical density of (\ref{twentysix}),
while the 2nd ensures the adiabatic condition (\ref{twentynine}) [7].

One clearly expects from (\ref{thirtyone}) a nonmonotonic suppression
of the solar $\nu_e$ flux as a function of neutrino energy, with the
maximum suppression occuring for some value of $E$ in between the two
limits.  As we shall see in the next section, this pattern seems to be
observed in the Gallium, Chlorine and the Water-Cherenkov experiments,
which are in increasing order of neutrino energy (see Fig. 2 and Table
II).  While the Gallium and the Water-Cherenkov experiments show a
suppression rate $\langle P_{\nu_e \rightarrow \nu_e}\rangle
\simeq 0.5$, the chlorine experiment shows a higher suppression rate 
of $\simeq 0.3$.

Finally, one sees from (\ref{thirty}) and (\ref{thirtyone}) that when
a solar neutrino experiment, corresponding to particular range of $E$,
finds a suppression rate for the solar $\nu_e$ flux, the result can
be cast into a contour plot in the $\delta m^2 - \sin^2 2\theta$
plane.  These contours are generally of triangular shape, as seen in
Fig. 3.  To understand why let us concentrate on the contour for the
Gallium experiments (SAGE and GALLEX), corresponding to $\langle
E\rangle \simeq 0.5$ MeV and $\langle P_{\nu_e \rightarrow
\nu_e}\rangle \simeq 0.5$.    The resonance
condition of (\ref{thirtyone}) fixes the $\delta m^2$ value, roughly
independent of $\sin^2 2\theta$, resulting in the horizontal side of
the triangle.  The 
adiabatic condition implies a minimum value of
the mixing angle, $\sin^2 2\theta \gsim 10^{-3}$.  At the
large angle end both the conditions of (\ref{thirtyone}) are
satisfied, so that there is complete MSW conversion between the two
neutrino species.  Consequently
\be
\langle P_{\nu_e \rightarrow \nu_e}\rangle = \sin^2 \theta, \ {\rm
i.e.} \ \sin^2 2\theta = 4 \langle P \rangle (1 - \langle P\rangle),
\label{thirtytwo}
\ee
which corresponds to the vertical side.  It extends downward by 3
orders of magnitude in $\delta m^2$, at which point the adiabatic
condition ceases to holds.  The third side corresponds to the
non-adiabatic solution to eq. (\ref{twentythree}).  Assuming a
constant density gradient, $d N_e/d\ell$, one gets [8]
\be
\langle P_{\nu_e \rightarrow \nu_e}\rangle \sim e^{-\pi \gamma_c/2}.
\label{thirtythree}
\ee
This corresponds to a contour of fixed $\delta m^2 \sin^2 2\theta$,
i.e. a diagonal line in the $\log \delta m^2 - \log \sin^2 2\theta$
plot (see eq. \ref{twentynine}).
\bigskip

\noindent {\Large{\bf Solar Neutrino Oscillation :}}
\medskip

The main sources of solar neutrinos are the three $pp$ chains of
nuclear reactions, taking place at the solar core, which convert
protons into $^4He$ ($\alpha$ particle).  They are
\beq
&{\rm (I)}& pp \rightarrow ^2H + e^+ + \nu_e, \ ^2H + p \rightarrow
^3He + \gamma, \ ^3He + ^3He \rightarrow ^4He + 2p; \ {\rm or}
\nonumber \\[2mm]
&{\rm (II)}& ^3He + ^4He \rightarrow ^7Be + \gamma, \ ^7Be + e^-
\rightarrow ^7Li + \nu_e, \ ^7Li + p \rightarrow 2 ^4He; \ {\rm or}
\nonumber \\[2mm]
&{\rm (III)}& ^7Be + p \rightarrow ^8B + \gamma, \ ^8B \rightarrow
^8Be^\star + e^+ + \nu_e, \ ^8Be^\star \rightarrow 2 ^4He.
\label{thirtyfour}
\eeq
While most of this conversion takes place by the straight path (I) a
small fraction takes place through a detour via $^7Be$ (II) and a
still smaller one through a longer detour via $^8B$ (III).  Thus the
$pp$ neutrino has the highest flux, followed by those of $^7Be$ and
$^8B$ neutrinos.  But their energies are in reverse order.  These are
shown in Fig. 2, along with the energy ranges of various solar
neutrino experiments [9].  The Gallium experiment gets largest
contribution from $pp$ neutrino (55\%), followed by the $^7Be$ (25\%)
and $^8B$ (10\%) neutrinos.  The chlorine experiment gets largest
contribution from $^8B$ neutrino (75\%) followed by the $^7Be$
(15\%).  The remainder in both cases come from the pep and the CNO
neutrinos.  The water Cherenkov experiment gets contribution only from
the $^8B$ neutrino.

The Gallium [10] and the Chlorine [11] experiments are based on the
reaction 
\be
\nu_e + ^{71}Ga \rightarrow e^- + ^{71} Ge,
\label{thirtyfive}
\ee
\be
\nu_e + ^{37} Cl \rightarrow e^- + ^{37} Ar.
\label{thirtysix}
\ee
The produced $^{71}Ge$ and $^{37}Ar$ are separated by radio chemical
method, from which the neutrino fluxes are estimated.  On the other
hand the water Cherenkov experiments are based on the charged current
interaction (\ref{twentyone}), where the outgoing electron is detected
by its Cherenkov radiation.  Thus it is a real time experiment.  It also
has directionality; the direction of the incoming neutrino can be
estimated from that of the outgoing electron.  Thus one can study
day-night (Zenith angle) variation of solar $\nu_e$ flux, which probes
the effect of its propagation through earth.  Similarly the energy
spectrum of the incoming neutrino can also be estimated from that of
the outgoing electron.  It may be noted here that this experiment also
gets a contribution from the Neutral Current reaction,
\be
\nu_{e,\mu} + e^- \ {\buildrel Z \over \longrightarrow} \ \nu_{e,\mu}
+ e^-;
\label{thirtyseven}
\ee
but at a reduced sensitivity of about 1/6th of the charged current
process. 

\begin{center}
Table II. The threshold energies of different solar neutrino 
experiments along \\ with the predicted and observed $\nu_e$ fluxes 
(in $10^6 \ cm^{-2} s^{-1}$ units) \\ for the
Kamiokande and event rates (in solar neutrino units) for the others.  \\ Their
ratio gives the survival probability shown in the last column.
\end{center}
\[
\begin{tabular}{|l|c|c|c|c|c|}
\hline
Expt. & Target & $E_{\rm th} ({\rm MeV})$ & Pred. & Obs. & $\langle
P_{\nu_e \rightarrow \nu_e}\rangle$ \\ 
\hline
GALLEX [10] & Gallium & 0.2 & $129^{+8}_{-6}$ & $78 \pm 8$ & $0.60 \pm .07$ \\
SAGE [10] & -''- & -''- & -''- & $67 \pm 8$ & $0.52 \pm .07$ \\
Homestake [11] & Chlorine & 0.8 & $7.7^{+1.2}_{-1.0}$ & $2.56 \pm .23$
& $0.33 \pm .05$ \\
Kamiokande [12] & Water & 7.5 & $5.15^{+1.0}_{-0.7}$ & $2.80 \pm .38$
& $0.54 \pm .07$ \\
Super-Kamiokande [12] & -''- & 6.5 & -''- & $2.44 \pm .10$ & $0.47 \pm
.08$ \\
\hline
\end{tabular}
\]

\noindent Table II lists the threshold energies of different solar
neutrino experiments along with the theoretically predicted and
experimentally observed results.  The ratio of the two
corresponds to the $\nu_e$ survival probability (supression rate)
$\langle P_{\nu_e 
\rightarrow \nu_e}\rangle$, shown in the last column.  As mentioned in
the last section, one can clearly see the nonmonotonic energy
dependence of this probability, as suggested by the MSW solution.  It
should be noted here that Kamiokande values for this quantity will go
down further by $\sim 0.1$ unit after taking account of the neutral
current interaction (\ref{thirtyseven}) effect. 

The MSW solutions to the observed suppression rates $\langle P_{\nu_e
\rightarrow \nu_e}\rangle$ for the 3 
sets of experiments are shown in Fig. 3 [13].  As discussed in the
last section, each solution corresponds to a triangular region in the
$\delta m^2 - \sin^2 2\theta$ plane.  The vertical position of each
triangle scales with the average neutrino energy of the corresponding
experiment (see eq. \ref{thirtyone}).  The overall solution to the
observed suppression rates corresponds to the overlap region among the
triangles, which are shown by the two shaded areas.  They represent
the small and large angle MSW solutions.  The additional constraints
coming from the Kamiokande energy spectrum and day-night effect are
also indicated.  However, they do not impinge upon the two allowed
regions.  Fig. 4 shows a more recent MSW analysis of the solar
neutrino data including the latest Super-Kamiokande results [14].
While one can have both the small and large angle solutions to the rates
and energy spectrum, only the small angle solution survives after
including the zenith angle distribution.

Finally Fig. 5 shows the corresponding vacuum oscillation solution to
the solar neutrino data [14].  One can reproduce the nonmonotonic
energy dependence of the suppression rate by assuming the half
wave-length $\lambda/2$ of eq. (\ref{eighteen}) to match with the
sun-earth distance, i.e. 
\be
1.25 E ({\rm MeV})/\delta m^2 ({\rm eV}^2) \simeq 10^{11} (m),
\label{thirtyeight}
\ee
for the middle energy range ($E \sim 5$ MeV).  Consequently one gets
the best solution for $\delta m^2 \sim 10^{-10} \ {\rm eV}^2$, and a
large mixing angle.  While the statistical significance of this
solution is as good as the small angle MSW, it seems less natural on
two counts -- the requirement of a large mixing angle and more
importantly the fine tuning of the sun-earth distance to match the
oscillation length for solar (MeV range) neutrino.  Interestingly in
this scenario one predicts seasonal variation, due to the eccentricity
of earth's orbit, which can be tested by future Super-Kamiokande
data.  In the absence of such measurement however the vacuum
oscillation solution appears less natural than the small angle MSW
solution. 
\bigskip

\noindent {\Large{\bf Atmospheric Neutrino Oscillation :}}
\medskip

The source of atmospheric neutrinos is the decay of $\pi^\pm$, which
are produced by the collision of cosmic rays with the atmosphere,
i.e. 
\beq
\pi^+ &\rightarrow& \mu^+ \nu_\mu, \ \mu^+ \rightarrow e^+ \nu_e
\bar\nu_\mu; \nonumber\\[2mm]
\pi^- &\rightarrow& \mu^- \bar\nu_\mu, \ \mu^- \rightarrow e^-
\bar\nu_e \nu_\mu.
\label{thirtynine}
\eeq
Thus one expects a ratio of
\be
R = {\nu_\mu + \bar\nu_\mu \over \nu_e + \bar\nu_e} = 2.
\label{fourty}
\ee
The observed ratio is significantly smaller -- close to 1 is some
situations.

In this case the (Super) Kamiokande is clearly the market leader [15],
although there is corroborative evidence from several other
experiments [16].  Interestingly the Kamiokande (as well as its super
version) started as Kamioka nucleon decay experiment and ended up as
Kamioka neutrino detection experiment.  In the process it showed that,
contrary to the conventional wisdom, yesterday's background can become
today's signal.  The atmospheric neutrinos have been long recognised
to constitute the irreducible background to proton decay.  Because of
this no earth-based experiment can probe proton life-time beyond
$10^{34}$ years, for which one has to go to the moon.  What the
(Super) Kamioka experiment did instead was to concentrate on the
study of this background.  And in the process they seem to have
discovered a result, which could be as significant as proton decay. 

The two main features of the SK data are as follows.

\begin{enumerate}
\item[{i)}] It shows a deficit in the $(\nu_\mu + \bar\nu_\mu)$ flux
while the $(\nu_e + \bar\nu_e)$ flux agrees with the prediction.  This
favours $\nu_\mu \rightarrow \nu_\tau$ oscillation over $\nu_\mu
\rightarrow \nu_e$, which is also supported by the CHOOZ reactor data
[17]. 
\item[{ii)}] The deficit is seen mainly for up-ward going $\nu_\mu$,
$\cos\Theta < 0$, where capital $\Theta$ denotes the zenith angle (to
avoid confusion with the mixing angle $\theta$).  Note that the
distance travelled by the up-ward going neutrino is related to earth's
diameter $d$ ($\sim 10^4$ Km) by
\be
\ell = d|\cos\Theta|,
\label{fourtyone}
\ee
while that of down-ward going ones is restricted to the atmospheric
depth ($\sim 10$ Km).  Hence this result suggest the oscillation
length $(\lambda/2)$ to be of similar order as $d$.
\end{enumerate}

Fig. 6 shows the zenith angle distribution of the electron and muon
neutrinos for different ranges of energy.  The theoretical
expectations without and with $(\nu_\mu \rightarrow \nu_\tau)$
oscillation are shown by hatched and solid lines respectively.  There
is no evidence for oscillation in the $\nu_e$ case.  But there is a
clear deficit of $\nu_\mu$ flux from the no oscillation prediction.
The lowest energy bin, corresponding to low oscillation length
$\lambda/2 \ll d$, shows a roughly constant deficit as expected from
(\ref{ninteen}) and (\ref{fourtyone}).  On the other hand multi-GeV
neutrinos correspond to higher oscillation length (\ref{eighteen}),
i.e ${\lambda \over 2} \sim d$.  In this case the deficit increases
steadily with the zenith angle, again in agreement with
(\ref{ninteen}) and (\ref{fourtyone}).

Fig. 7 combines the data points from different zenith angles
(\ref{fourtyone}) and energy to give the distribution in the ratio
$\ell/E$.  This is the appropriate quantity for studying the
oscillation phenomenon via (\ref{seventeen}).  One sees a large
decrease in the $\nu_\mu$ flux, relative to the Monte Carlo
prediction, by almost a factor of 2 as $\ell/E
\rightarrow 10^3$ Km/GeV.  It corresponds to a large mixing angle
$(\sin^2 2\theta \sim 1)$ and a $\delta m^2 \sim 10^{-3} \ {\rm
eV}^2$. Indeed the best solution represented by the dashed line,
corresponds to $\sin^2 2\theta = 1$ and $\delta m^2 = 2.2 \times
10^{-3} \ {\rm eV}^2$. 

Finally Fig. 8 shows the allowed region of the vacuum oscillation
solution to the Super-Kamiokande data in the $\sin^2 2\theta$ and
$\delta m^2$.  The corresponding 90\% CL contour of the earlier
Kamiokande data is also shown for comparison.  Although the latter is
slightly higher in $\delta m^2$, the two agree within $1.5\sigma$.  Of
course the combined 90\% CL contour will be close to the SK contour
because of its higher statistical significance, i.e.
\be
\sin^2 2\theta_{\mu\tau} > 0.82 (\theta_{\mu\tau} = 45 \pm
13^0),
\label{fourtytwo}
\ee
and
\be
\delta m^2 = (0.5 - 6)10^{-3} \ {\rm eV}^2.
\label{fourtythree}
\ee
\bigskip

\noindent {\Large{\bf Reconciling Large Mixing with Hierarchical
Masses :}}
\medskip

Thus the atmospheric neutrino oscillation data implies nearly maximal
mixing between $\nu_\mu$ and $\nu_\tau$ (\ref{fourtytwo}).  This would
normally suggest a degenerate pair of neutrinos, with a small mass
difference given by (\ref{fourtythree}).  For example a Dirac
mass-matrix for the $\nu_\mu - \nu_\tau$ pair would correspond to
degenerate masses and maximal mixing, i.e.
\be
\left(\matrix{0 & M \cr M & 0}\right) \rightarrow \left(\matrix{M & 0
\cr 0 & -M}\right), \theta = 45^\circ.
\label{fourtyfour}
\ee
However the favoured solution to the solar neutrino oscillation data
(Fig. 4) would then require the $\nu_e$ to have even a more precise
mass degeneracy with one of these states with 
\be
\delta m^2 = (0.5 - 1) 10^{-5} \ {\rm eV}^2
\label{fourtyfive}
\ee
and 
\be
\sin^2 2\theta_e = 10^{-3} - 10^{-2} \left(\sin \theta_e = {1\over50} -
{1\over20}\right). 
\label{fourtysix}
\ee
Such a degeneracy would of course be totally unexpected.  It is
therefore more natural to consider the alternative of hierarchical
neutrino masses instead of degenerate ones.  It implies that the two
larger mass eigen values correspond to the square-roots of
eq. (\ref{fourtythree}) and (\ref{fourtyfive}),
\be
m_1 \sim 0.05 \ {\rm eV}, \ m_2 \sim 0.003 \ {\rm eV},
\label{fourtyseven}
\ee
with $m_3 \ll m_2$.  Indeed as far as the atmospheric and
solar neutrino oscillation phenomena are concerned one can take $m_3$
to be exactly $0$.  Note that the $m_1$ and $m_2$ mass eigen states
correspond to large admixtures of $\nu_\mu$ and $\nu_\tau$
(\ref{fourtytwo}), with a small $\nu_e$ component corresponding to
(\ref{fourtysix}). 

There is a broad consensus in the current literature in favour of this
second alternative.  Indeed much of it is devoted to exploring models
for reconciling hierarchical masses with at least one large mixing
angle.  I shall conclude by briefly discussing a minimal see-saw model
where the two can be naturally reconciled. 

Recall that the canonical see-saw model [2] is based on a $U(1)$
extension of the SM gauge group, corresponding to the gauge charge
$B-L$.  The latter treats the three neutrino flavours identically,
while the atmospheric and solar neutrino oscillation data seem to
distinguish the $\nu_e$ from $\nu_\mu$ and $\nu_\tau$.  Thus to
account for these oscillation data one needs to consider a variation
of this see-saw model, where the above distinction between the
neutrino flavours is incorporated into the choice of the $U(1)$ gauge
charge.  Two such variations were presented by us in [18] and [19].  I
shall concentrate on the latter, because it is more economical.  It
corresponds to the $U(1)$ gauge charge
\be
Y' = B - {3\over2} (L_\mu + L_\tau).
\label{fourtyeight}
\ee
The requirement of anomaly cancellation implies two right-handed
singlet neutrinos $N_{1,2}$ with $Y' = -3/2$ to match the two doublet
ones, carrying this gauge charge.

The minimal Higgs sector consists of
\be
\left(\matrix{\phi^+ \cr \phi^0}\right)_{Y' = 0} \ \& \ \chi^0_{Y'=3} 
\label{fourtynine}
\ee
as in the canonical see-saw model.  The latter acquires a large vev at
the $Y'$ breaking scale giving large Majorana masses to $N_{1,2}$,
i.e. 
\be
M_{1,2} \sim \langle \chi \rangle.
\label{fifty}
\ee
The $\phi$ couples these right-handed singlet neutrinos to $\nu_\mu$
and $\nu_\tau$ giving them Dirac masses,
\be
f \phi \bar\nu_{\mu,\tau} N_{1,2} \rightarrow f\langle \phi \rangle
\bar \nu_{\mu,\tau} N_{1,2},
\label{fiftyone}
\ee
while there is no such coupling to $\nu_e$.  Thus it implies two
non-zero mass eigen states of the SM neutrinos, corresponding to large
admixtures of $\nu_\mu$ and $\nu_\tau$ but no $\nu_e$ component.  In
order to introduce a small mixing of $\nu_e$ with these states, as
required by the solar neutrino oscillation data, we expand the Higgs
sector by an additional doublet and a singlet,
\be
\left(\matrix{\eta^+ \cr \eta_0}\right)_{Y' = -3/2} \ \& \ \zeta^0_{Y' = -3/2}.
\label{fiftytwo}
\ee
The doublet introduces a small Dirac coupling of $\nu_e$ with $N_{1,2}$
via
\be
f \eta \bar\nu_e N_{1,2} \rightarrow f\langle \eta \rangle \bar\nu_e
N_{1,2}. 
\label{fiftythree}
\ee
The singlet $\zeta^0$ is required to avoid an unwanted
pseudo-Goldstone boson.  This comes about because there are three
global $U(1)$ symmetries, corresponding to rotating the phases of
$\phi,\eta$ and $\chi^0$ in the Higgs potential, while only two local
$U(1)$ symmetries get broken.  The addition of the singlet $\zeta^0$
introduces two more terms in the Higgs potential, $\eta^\dagger 
\phi \zeta^0$ and $\chi^0 \zeta^0 \zeta^0$, so that the extra global
symmetry is avoided.  

The doublet $\eta$ can acquire a small but nonzero vev at the $SU(2)$
breaking scale, which can be estimated from the relevant part of the
Higgs potential,
\be
m^2_\eta \eta^+ \eta + \lambda(\eta^\dagger \eta) (\chi^\dagger \chi)
+ \lambda'(\eta^\dagger \eta) (\zeta^\dagger \zeta) - \mu
\eta^\dagger \phi \zeta.
\label{fiftyfour}
\ee
Although $m^2_\eta$ is positive, after minimisation of the potential
we find this field has acquired a small vev,
\be
\langle \eta \rangle = \mu \langle \phi \rangle \langle \zeta 
\rangle / M^2_\eta,
\label{fiftyfive}
\ee
where $M^2_\eta = m^2_\eta + \lambda \langle \chi \rangle^2 + \lambda'
(\zeta\rangle^2$ represents the physical mass of $\eta$ and $\mu
\lsim \langle \zeta\rangle \sim \langle \chi \rangle$.  Thus a
reasonable choice of $M_\eta \sim 5 \langle \zeta\rangle$ implies 
\be
\langle \eta \rangle /\langle \phi \rangle \sim 1/25,
\label{fiftysix}
\ee
which would correspond to the required mixing angle for $\nu_e$
(\ref{fourtysix}). 

Let us write down the full $5 \times 5$ mass matrix in the flavour
basis of $\nu_{e,\mu,\tau}$, corresponding to the mass basis of the
charged leptons.  Since the latter do not interact with the singlet
neutrinos $N_{1,2}$ they can be simultaneously diagonalised in the
same basis.  Thus we have the symmetric mass matrix 
\be
\left(\matrix{0 & \cdot & \cdot & \cdot & \cdot \cr 0 & 0 & \cdot &
\cdot & \cdot \cr 0 & 0 & 0 & \cdot & \cdot \cr f^1_e \langle \eta
\rangle & f^1_\mu \langle \phi \rangle & f^1_\tau \langle \phi \rangle &
M_1 & \cdot \cr f^2_e \langle \eta \rangle & f^2_\mu \langle \phi
\rangle & f^2_\tau \langle \phi \rangle & 0 & M_2}\right).
\label{fiftyseven}
\ee
We assume the Higgs Yukawa couplings to be of similar magnitude, which
means that a mass-matrix arising from a single Higgs vev has a
democratic texture.  On the other hand, in analogy with the charged
leptons, it is reasonable to assume a mild hierarchy for the $N_{1,2}$
mass eigen values, i.e.
\be
M_1/M_2 \sim 1/20.
\label{fiftyeight}
\ee
There is of course no conflict between the two assumptions; rather
they are closely connected.  Democratic mass matrices naturally lead
to large cancellations in the determinant, which are required for
hierarchical mass eigen values. 

One can write down the $3 \times 3$ mass-matrix for the SM neutrinos
from (\ref{fiftyseven}) using the see-saw formula [19].  Instead of
doing that, we will simply read off the rough magnitudes of the masses
and mixings from the matrix elements of (\ref{fiftyseven}).  We get
\beq
\tan\theta_{\mu\tau} &\sim& f^1_\mu/f^1_\tau \sim 1, \nonumber \\[2mm] 
\sin\theta_e &\sim& \langle \eta \rangle/\langle\phi\rangle \sim 1/25,
\nonumber \\[2mm] m_2/m_1 &\sim& M_1/M_2 \sim 1/20,
\label{fiftynine}
\eeq
which are in good agreement with the experimental value of
(\ref{fourtytwo}), (\ref{fourtysix}) and (\ref{fourtyseven}).  Finally
the scale of the $U(1)_{Y'}$ symmetry breaking can be estimated from
the see-saw formula,
\beq
M_2 = f^2 \langle \phi \rangle^2/m_2 &=& f^2 10^{16} \ {\rm GeV}
\nonumber \\[2mm] &=& 10^{12} - 10^{16} \ {\rm GeV},
\label{sixty}
\eeq
depending on whether we take the $\phi$ Yukawa coupling as $f \sim
10^{-2}$ in analogy with the $\tau$ lepton or $\sim 1$ in analogy with
top quark.  Thus for a $U(1)_{Y'}$ symmetry breaking scale of $10^{12}
- 10^{16}$ GeV the model can naturally account for the large (small)
mixing solutions to the atmospheric (solar) neutrino oscillation
data. 

Let me mention here that there is one more evidence of neutrino
oscillation, from the Los Alamos accelerator neutrino experiment [20],
which is far more controversial however than the solar and atmospheric
neutrino results.  If confirmed, it will be hard to explain all the
three oscillation results within the framework of three light
neutrinos. 

I thank Girish Ogale for typing the manuscript and Rajan Pawar for
drafting the 1st figure.
\bigskip

\noindent {\Large{\bf References}}
\medskip

\begin{enumerate}
\item[{1.}] M. Gell-Mann, P. Ramond and R. Slansky, in Supergravity,
Proceedings of the Workshop, Stony Brook, New York, 1979
(North-Holland, Amsterdam); T. Yanagida, in Proceedings of the
Workshop on Unified Theories and Baryon Number in the Universe,
Tsukuba, Japan (KEK Report No. 79-18) 1979.
\item[{2.}] R.E. Marshak and R.N. Mohapatra, Phys. Lett. B91, 222
(1980).
\item[{3.}] B. Pontecorvo, JETP 33, 549 (1957); 34, 247 (1958); 53,
1717 (1967).
\item[{4.}] S.P. Mikheyev and A.Y. Smirnov, Yad. Fiz. 42, 1441 (1985);
Nuovo Cim. 9C, 17 (1986).
\item[{5.}]  L. Wolfenstein, Phys. Rev. D17, 2369 (1978).
\item[{6.}] H. Bethe, Phys. Rev. Lett. 56, 1305 (1986).
\item[{7.}] See e.g. K. Whisnant, Ashland 1987, Proceedings of
Neutrino Masses and Neutrino Astrophysics.
\item[{8.}] S.J. Parke, Phys. Rev. Lett. 57, 1275 (1986); S.J. Parke
and T.P. Walker, Phys. Rev. Lett. 57, 2322 (1986).
\item[{9.}] S.T. Petcov, hep-ph/9806466; J.N. Bahcall, Neutrino
Astrophysics, Cambridge University Press, Cambridge, 1989.
\item[{10.}] GALLEX Collaboration: W. Hampel et. al.,
Phys. Lett. B388, 364 (1996); SAGE Collaboration: V. Garvin et. al.,
Neutrino-98, Takayama, Japan (1998).
\item[{11.}] Homestake Expt.: B.T. Cleveland et. al.,
Astrophys. J. 496, 505 (1998); R. Davis, Prog. Part. Nucl. Phys. 32,
13 (1994).
\item[{12.}] Kamiokande Collaboration: K.S. Hirata et. al.,
Phys. Rev. Lett. 77, 1683 (1996); Super-Kamiokande Collaboration:
Y. Suzuki, Neutrino-98, Takayama, Japan (1998).
\item[{13.}] N. Hata and P.G. Langacker, Phys. Rev. D56, 6107 (1997).
\item[{14.}] J.N. Bahcall, P.J. Krastev and A.Y. Smirnov,
Phys. Rev. D58, 096016 (1998).
\item[{15.}] Kamiokande Collaboration: Y. Fukuda et. al.,
Phys. Lett. B335, 237 (1994); Super-Kamiokande Collaboration:
Phys. Rev. Lett. 81, 1562 (1998).
\item[{16.}] IMB Collaboration: R. Becker-Szendy et. al.,
Nucl. Phys. (Proc. Suppl.) 38, 331 (1995); Soudan-2 Expt.:
W.W.M. Allison et. al., hep-ex/9901024.
\item[{17.}] CHOOZ Collaboration: M. Appollonio et. al.,
Phys. Lett. B420, 397 (1998).
\item[{18.}] Ernest Ma, D.P. Roy and Utpal Sarkar, Phys. Lett. B444,
391 (1998).
\item[{19.}] Ernest Ma and D.P. Roy, hep-ph/9811266 (to be published
in Phys. Rev. D).
\item[{20.}] LSND Collaboration: A. Athanassopoalos et. al.,
nucl-ex/9706006. 
\end{enumerate}

\newpage

\begin{figure}[htb]
\begin{center}
\leavevmode
\hbox{%
\epsfxsize=5in
\epsffile{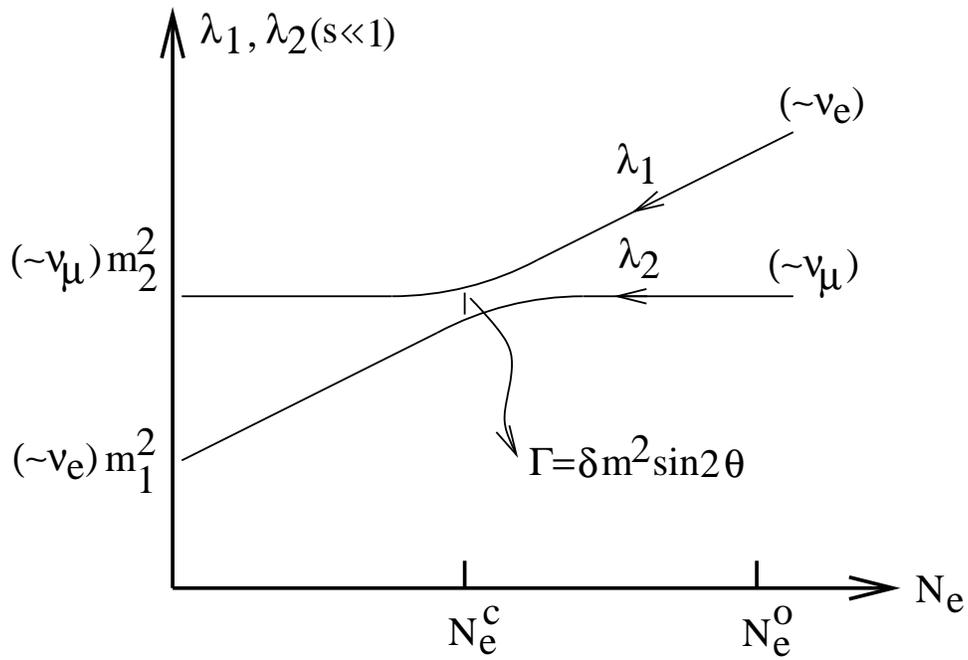}}
\caption{Schematic diagram of the energy eigenvalues of
$\nu_e$ and $\nu_\mu$ as functions of electron density; $N^0_e$
denotes the electron density at the solar
core and $N^c_e$ the critical density where the two energy levels
cross [6,7].} 
\label{fig:nufig1}
\end{center}
\end{figure}

\begin{figure}[htb]
\begin{center}
\leavevmode
\hbox{%
\epsfxsize=4.5in
\epsffile{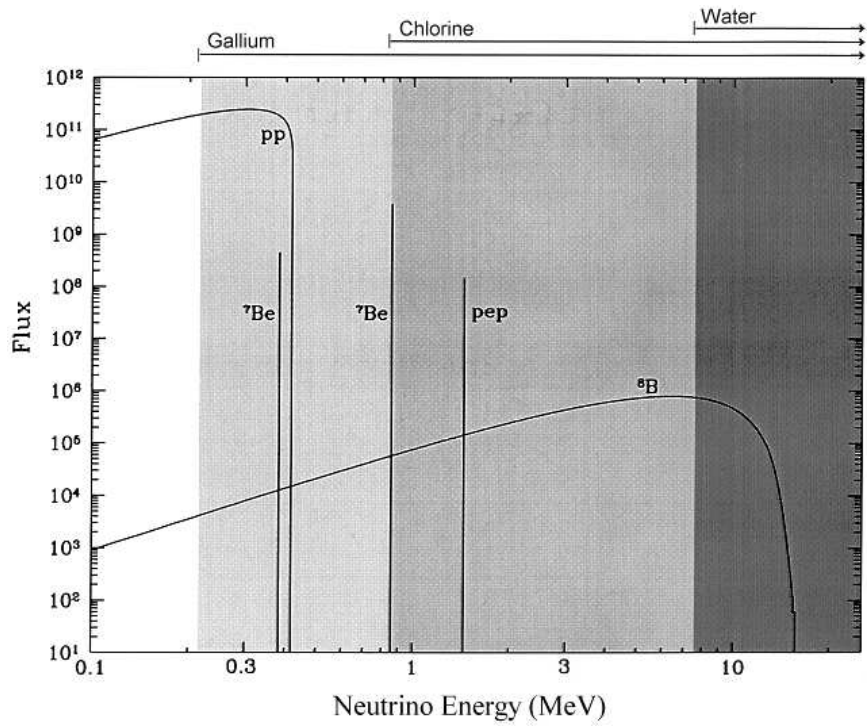}}
\caption{The spectra of the $pp$, $pep$, $^7Be$ and $^8B$
neutrinos are shown along with the energy ranges of different solar
neutrino experiments [9].}
\label{fig:nufig2}
\end{center}
\end{figure}

\newpage

\begin{figure}[htb]
\begin{center}
\leavevmode
\hbox{%
\epsfxsize=5in
\epsffile{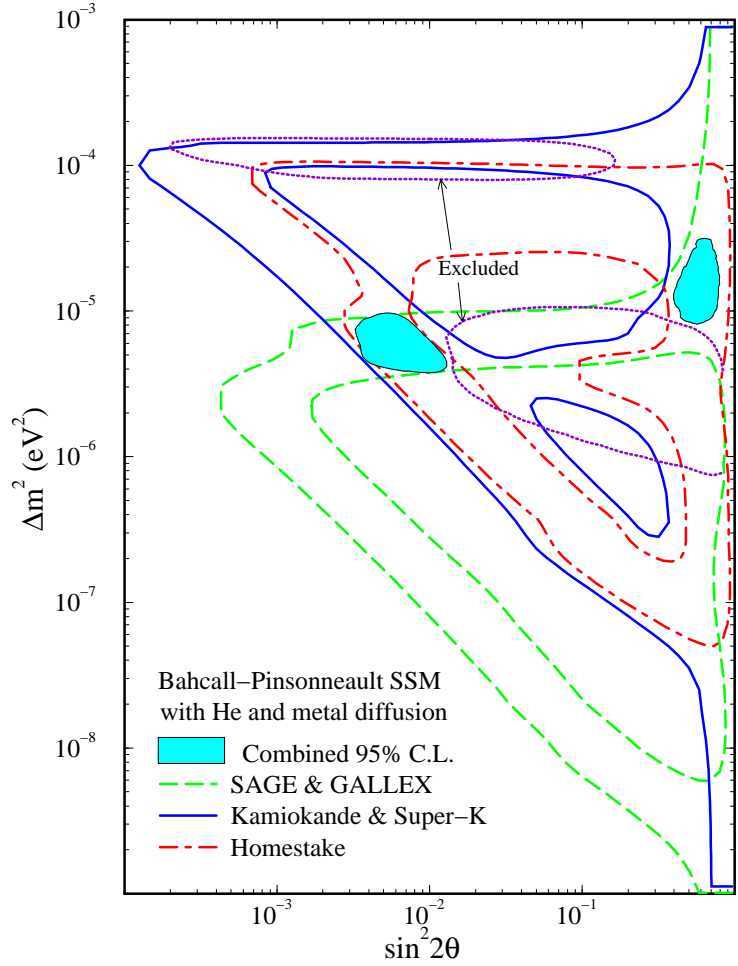}}
\caption{The 95\% CL contours of MSW solution for the
suppression rates observed by different solar neutrino experiments
[13].  Also shown are the regions excluded by the Kamiokande
energy spectrum and day-night asymmetry.}
\label{fig:nufig3}
\end{center}
\end{figure}

\newpage

\begin{figure}[htb]
\begin{center}
\leavevmode
\hbox{%
\epsfxsize=3in
\epsffile{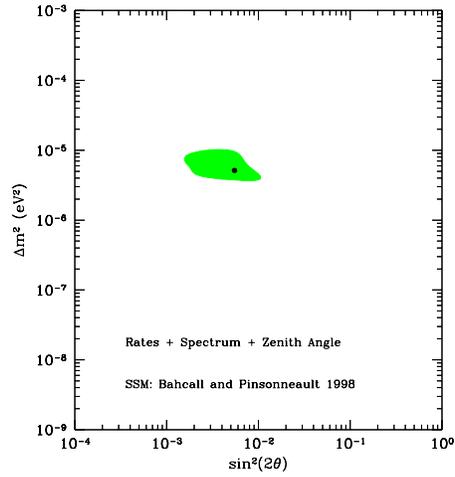}}
\caption{The MSW solution to the
combined solar neutrino data on suppression rates along with the
Super-Kamiokande energy spectrum and zenith angle distribution.  The
contour is drawn at 99\% CL [14].}
\label{fig:nufig4}
\end{center}
\end{figure}

\begin{figure}[htb]
\begin{center}
\leavevmode
\hbox{%
\epsfxsize=3in
\epsffile{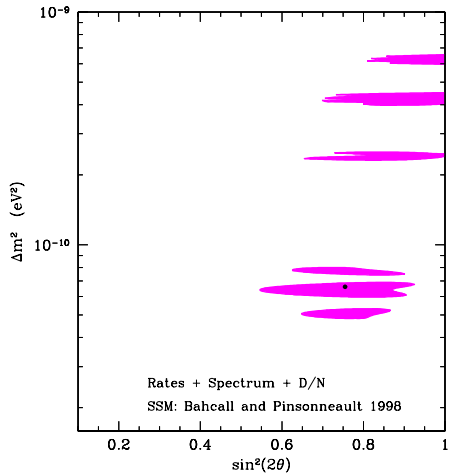}}
\caption{The vacuum oscillation solution to the combined solar
neutrino data on suppression rates along with the Super-Kamiokande
energy spectrum and day-night asymmetry.  The contours are drawn at
99\% CL [14].}
\label{fig:nufig5}
\end{center}
\end{figure}

\newpage

\begin{figure}[htb]
\begin{center}
\leavevmode
\hbox{%
\epsfxsize=5in
\epsffile{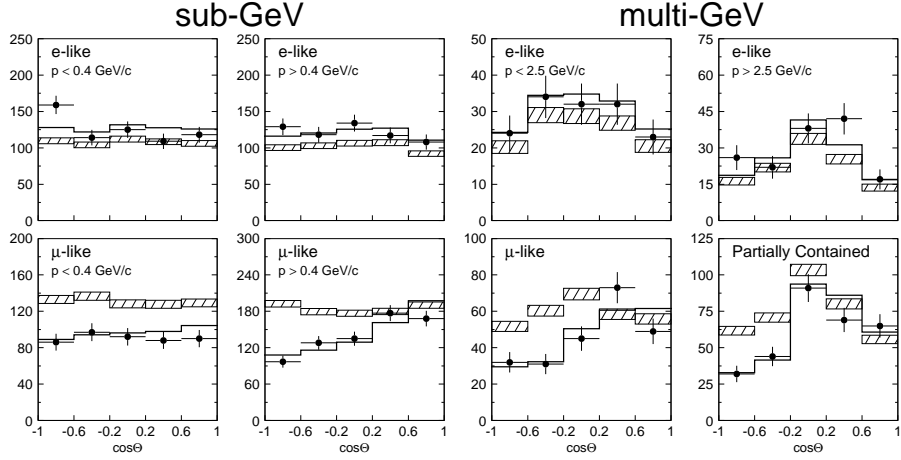}}
\caption{Zenith angle distribution of atmospheric $\nu_e$
$(\bar\nu_e)$ and $\nu_\mu$ $(\bar\nu_\mu)$ events from
Super-Kaniokande for different energy ranges.  The partially contained
$\nu_\mu$ $(\bar\nu_\mu)$ events roughly correspond to $p > 10$
GeV/c.  The theoretical prediction with and without $\nu_\mu
\rightarrow \nu_\tau$ oscillation are shown by solid and hatched lines
respectively [15].}
\label{fig:nufig6}
\end{center}
\end{figure}

\begin{figure}[htb]
\begin{center}
\leavevmode
\hbox{%
\epsfxsize=3in
\epsffile{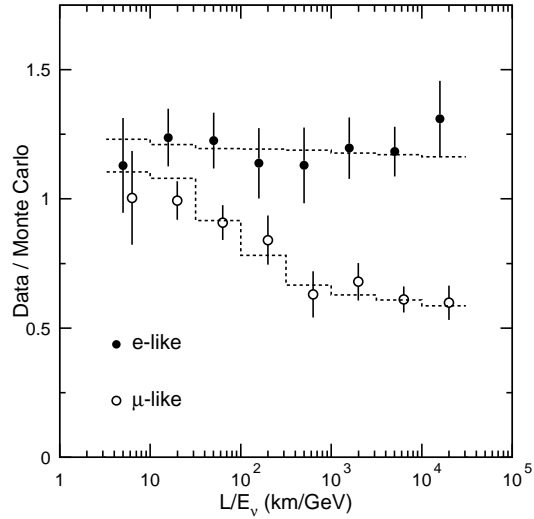}}
\caption{The survival probabilities $P_{\nu_e \rightarrow
\nu_e}$ and $P_{\nu_\mu \rightarrow \nu_\mu}$ observed by the
Super-Kamiokande atmospheric neutrino experiment are shown against the
reconstructed $\ell/E_\nu$.  The theoretical prediction with the
without $\nu_\mu \rightarrow \nu_\tau$ oscillation are shown by the
lower and upper lines respectively [15].}
\label{fig:nufig7}
\end{center}
\end{figure}

\begin{figure}[htb]
\begin{center}
\leavevmode
\hbox{%
\epsfxsize=5in
\epsffile{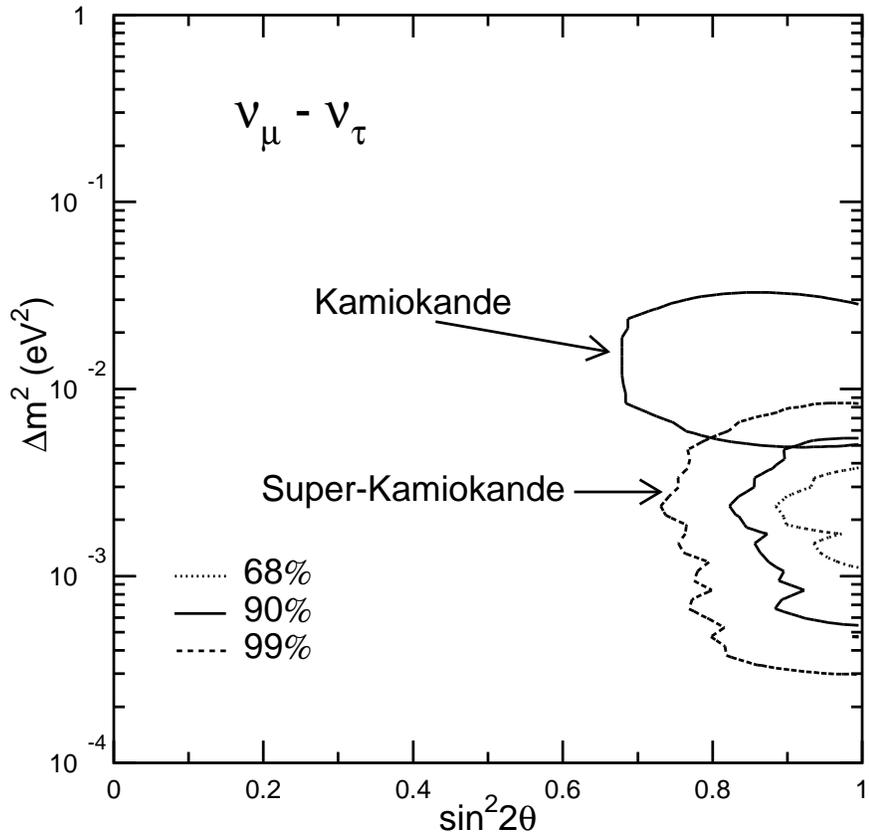}}
\caption{The $\nu_\mu \rightarrow \nu_\tau$ oscillation
solutions to the Super-Kamiokande atmospheric neutrino data are shown
at 68\%, 90\% and 99\% CL.  The corresponding 90\% CL contour for the
Kamiokande data is also shown for comparison [15].}
\label{fig:nufig8}
\end{center}
\end{figure}

\end{document}